\documentclass[twocolumn,amsmath,amssymb,aps]{revtex4}

\usepackage{times}
\usepackage{amsmath}
\usepackage{amssymb}
\usepackage[]{graphicx}
\usepackage{bm}
\usepackage{color}
\usepackage{soul}
\usepackage{float}
\usepackage{natbib}
\usepackage{latexsym}
\usepackage{multirow}

\begin{document}

\title{Effects of Fluctuating Energy Input on the Small Scales in Turbulence}

\author{Chen-Chi Chien, Daniel B. Blum, Greg A. Voth$^\ast$\\
Department of Physics, Wesleyan University,\\ Middletown, Connecticut 06459, USA\\
$^\ast$ Correspondence to: gvoth@wesleyan.edu
}

\date{\today}

\begin{abstract}

In the standard cascade picture of 3D turbulent fluid flows, energy is input at a constant rate at large scales.  Energy is then transferred to smaller scales by an intermittent process that has been the focus of a vast literature.  However, the energy input at large scales is not constant in most real turbulent flows.  We explore the signatures of these fluctuations of large scale energy input on small scale turbulence statistics.   Measurements were made in a flow between oscillating grids, with $R_{\lambda}$ up to 271, in which temporal variations in the large scale energy input can be introduced by modulating the oscillating grid frequency.  We find that the Kolmogorov constant for second order longitudinal structure functions depends on the magnitude of the fluctuations in the large scale energy input.  We can quantitatively predict the measured change with a model  based on Kolmogorov's refined similarity theory.    The effects of fluctuations of the energy input can also be observed using structure functions conditioned on the instantaneous large scale velocity.   A linear parameterization using the curvature of the iconditional structure functions provides a fairly good match with the measured changes in the Kolmogorov constant.  Conditional structure functions are found to provide a more sensitive measure of the presence of fluctuations in the  large scale energy input than inertial range scaling coefficients.  \end{abstract}
\maketitle

\section{\label{sec:intro}Introduction}

One of the earliest recognitions of the importance of fluctuations in the energy dissipation rate in turbulence can be found in a footnote by Landau in the textbook on fluid mechanics \cite{Landau:1959}.   The footnote explains that universal formulas for the small scales of structure functions do not exist because the energy dissipation rate will fluctuate on long time scales, and these fluctuations will be different in different flows. Frisch \cite{Frisch:1995} provides an extended discussion of the footnote.    In the refined similarity theory by Kolmogorov~\cite{Kolmogorov:1962} and Obukhov~\cite{Obukhov:1962}, this insight on universality is extended to include fluctuations that result from the random character of the transfer of energy between scales, which is often called internal intermittency.  Kolmogorov~\cite{Kolmogorov:1962} gives Landau credit for recognizing the importance of internal intermittency.  However, this credit seems to be somewhat misplaced since  the available published text by Landau observes only that large scale fluctuations in the energy dissipation will destroy universality of small scales \cite{Frisch:1995, Mouri:2006}.  During the intensive effort  to understand internal intermittency over the past 50 years, the direct application of Landau's insight about the importance of large scale fluctuations has often been obscured.  

The refined similarity theory by Obukhov~\cite{Obukhov:1962} and Kolmogorov~\cite{Kolmogorov:1962} proposed that in the inertial range the moments of velocity differences between two points are universal functions when they are conditioned on the  locally averaged value of the energy dissipation rate, $\varepsilon_r$, defined as the instantaneous energy dissipation rate averaged over a sphere of radius $r$.   For simplicity we will consider the longitudinal component of the velocity differences, $\Delta_r u$. The conditional moments are
  
\begin{equation}
\label{eqn:Delta_r}
\langle(\Delta_r u)^p|\varepsilon_r \rangle = C_p(\varepsilon_r r)^{p/3},
\end{equation}
where $C_p$ are universal constants~\citep{Pope:2000}.   Averaging this expression over a distribution of $\varepsilon_r$ yields
\begin{equation}
  \langle(\Delta_ru)^p\rangle =  C_p \langle\varepsilon_r^{p/3} \rangle r^{p/3} =  C_p \frac{\langle\varepsilon_r^{p/3} \rangle}{\varepsilon^{p/3}} (\varepsilon r)^{p/3},
 \label{refined_avg}
 \end{equation}
 where $\varepsilon=\langle \varepsilon_r \rangle$  is the mean energy dissipation rate.  Since the moments of $\varepsilon_r$ depend on $r$, this means that the inertial range scaling law is modified by internal intermittency.  Kolmogorov proposed that the fluctuations of $\varepsilon_r$ could be described with a power law scaling 
\begin{equation}
\frac{\langle \varepsilon_r^p \rangle}{ \varepsilon ^p} \propto \left(\frac{L}{r}\right)^{\xi_p}.
\end{equation}
where $L$ is a length characterizing the energy input scale.  In Kolmogorov (1962)\cite{Kolmogorov:1962}, a log-normal model was used to relate $\xi_p$ for all $p$ to $\xi_2=\mu$, which is commonly called the intermittency exponent.   An extensive literature has explored the $r$ dependence of statistics of $\varepsilon_r$ in order to understand anomalous scaling exponents in the inertial range ~\cite{Sreenivasan:1997}. 

However, the effects of fluctuations in the energy dissipation rate due to the large scales has been given much less attention, even though this is the direct application of Landau's original comment.   Kolmogorov did state that the coefficients in the scaling law should not be universal, presumably because he recognized that large scale fluctuations would not be universal \cite{Kolmogorov:1962}.  
Monin and Yaglom \cite{Monin:1971} provide a simple model at the beginning of their section titled ``Refined Treatment of the Local Structure of Turbulence, taking into account fluctuations in the dissipation rate''.  An extended presentation of this model is in the textbook by Davidson~\cite{Davidson:2004}.  They consider averaging together equal numbers of samples from two different turbulent states: state 1 with energy dissipation rate $\varepsilon_1=(1+\gamma)\langle \varepsilon \rangle$ and another state 2 with $\varepsilon_2=(1-\gamma)\langle \varepsilon \rangle$.  Here $\langle \varepsilon \rangle$ is the mean energy dissipation rate and $\gamma$ is a measure of the difference in energy dissipation between the two states.   Then equation~(\ref{refined_avg}) implies that a measured second order structure function in the inertial range averaged over equal contributions from each state would be
\begin{equation}
  \langle(\Delta_ru)^2\rangle =  \frac{C_2}{2} \left[ (1+\gamma)^{2/3} + (1-\gamma)^{2/3} \right]  (\varepsilon r)^{2/3}.
 \label{MY_model}
 \end{equation}
 So the large scale fluctuations in the energy dissipation are predicted to change the coefficient of the inertial range scaling law without changing the power law scaling.   In this model, $\gamma$ must be less than or equal to one, so the coefficient of the second order structure function can decrease to as low as $C_2/2^{1/3} \approx 0.794 \: C_2$ for the case $\gamma=1$ where there is no energy injection in state 2. 
 
 This model is easily extended to the case where samples are included from state 1 with probability $\beta$ and from state 2 with probability $1-\beta$.    Now the energy dissipation rates are $\varepsilon_1=(1+(1-\beta)\gamma/\beta)\langle \varepsilon \rangle$ and 
$\varepsilon_2=(1-\gamma)\langle \varepsilon \rangle$.   For this extended model, the measured structure function of order $p$ would be 
 \begin{equation}
  \langle(\Delta_ru)^p\rangle=  \kappa(\beta,\gamma)  \; C_p  \left(\langle \varepsilon \rangle r \right)^{p/3}.
 \label{MY_model_duty}
 \end{equation} 
 where the correction factor of the coefficient is 
  \begin{equation}
 \kappa(\beta,\gamma)=\left[ \beta \left(1+\frac{1-\beta}{\beta} \gamma \right)^{p/3} + (1-\beta)\left(1-\gamma\right)^{p/3}\right].
 \label{Eq:correction_factor}
 \end{equation}
 In the limiting case $\gamma=1$ and $\beta \rightarrow 0$, the coefficient for $p=2$ goes to zero, and the coefficients for $p>3$ go to infinity, so the effects of large scale fluctuations on the small scale statistics can be very large.   In this limiting case, the flow consists of brief pulses of large energy input between long periods of no energy input.  

\begin{figure}
\begin{center}
\includegraphics[width=3.3in]{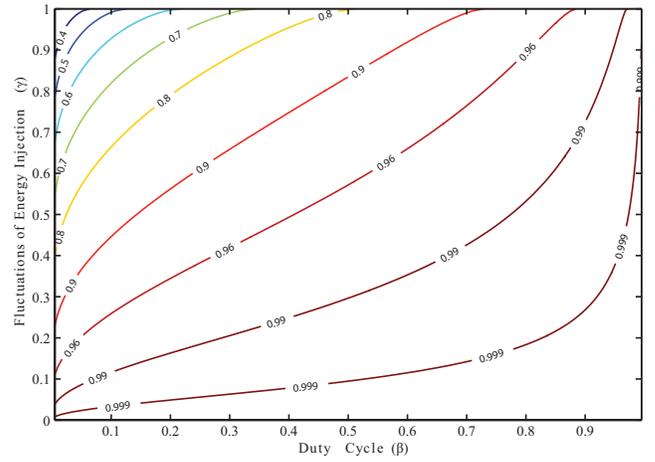}
\caption[The correction factor $ \kappa $ for $p=2$]{\label{fig:contour}   Contour plot of the correction factor $ \kappa $ from equation~(\ref{Eq:correction_factor}) for $p=2$, showing the change in coefficients in the inertial range the scaling law as a function
of the amplitude of fluctuations in the energy input $\gamma$, and the time spent in the high energy input state, or duty cycle, $\beta$.  }
 \end{center}
 \end{figure}

 In both Monin and Yaglom \cite{Monin:1971} and Davidson \cite{Davidson:2004}, the presentation of the model in equation (\ref{MY_model}) is followed with the observation that in typical situations this effect is not large.   
Figure~\ref{fig:contour} shows a contour plot of the correction factor for $p=2$ in equation~\ref{Eq:correction_factor} as a function of the fluctuations in the energy input, $\gamma$, and the fraction of the time spent in the high energy input state, or duty cycle, $\beta$.   The observation that the correction is not large in most cases is justified since the correction is less than 2.4\% for half of the parameter space for $p=2$.  However, the correction can be very large in some flows.  There is always a divergence for  $\gamma=$1 and $\beta \rightarrow $0, and for large $p$, the correction is larger.  Although this two state model is a simple idealization, we will show that it provides a reasonably good description of some of our data.   

In real flows, the energy dissipation rate and $\varepsilon_r$ have continuous distributions.  In the continuous case,  equation~(\ref{refined_avg}) can be used to predict the behavior of structure functions, but there are now contributions to the distribution of $\varepsilon_r$ from both internal intermittency and fluctuations in the energy input. In particular, $\varepsilon_r$ for $r \ge L$ has a distribution which is determined not by cascade processes but by the mechanisms creating the turbulence.  In cases where internal intermittency can be ignored, we can estimate the fluctuations in the energy input and predict the coefficients of scaling law.  
If the mean square velocity, $3U^2=\langle u_i u_i \rangle$ and the integral length scale, $L$, are defined using ensemble averages, then they can be considered to be time dependent.  In this case, $\varepsilon \propto U^3/L$ provides an estimate of the instantaneous energy dissipation rate.  If time averages of this dissipation rate are then used in equation~(\ref{refined_avg}) we obtain
\begin{equation}
\langle(\Delta_r u)^p\rangle = C_p \frac{\langle(U^3/L)^{p/3} \rangle}{\langle U^3/L\rangle^{p/3}} (\varepsilon r)^{p/3}.  
 \label{refined_model_u}
 \end{equation}
 In our flow, where $L$ has a weak dependence on the variations in the energy input, this simplifies to 
\begin{equation}
\langle(\Delta_ru)^p\rangle = C_p \frac{\langle U^p \rangle}{\langle U^3 \rangle^{p/3}} (\varepsilon r)^{p/3}.
 \label{refined_model}
 \end{equation}
  If internal intermittency is also important, then the two effects may be combined as
 \begin{equation}
  \langle(\Delta_ru)^p\rangle =   C'_p  \frac{\langle(U^3/L)^{p/3} \rangle}{\langle U^3/L\rangle^{p/3}}   \left(\frac{L}{r}\right)^{\xi_p} (\varepsilon r)^{p/3}.
 \label{refined_intermittent}
 \end{equation}

It is important to determine the size of the effects of fluctuations in the large scale energy input in real turbulent flows.  Surprisingly, there are no published results that we know of that document a dependence of coefficients of inertial range scaling laws for structure functions on systematic changes in the large scales of the flow.  A compilation of experimental~\cite{Sreenivasan:1995} and simulation~\cite{Donzis:2010} results have given credence to the notion that the second order coefficients are close enough to  independent of the flow that they can be treated as universal constants.  At least three experimental studies have explored fluctuations in the large scale energy input in detail.  Praskovsky et al. \cite{Praskovsky:1993}  study two high Reynolds number flows, a mixing layer and a return channel.  They find a conditional dependence of the second order structure functions on the instantaneous velocity and connect this with spatial and temporal variability of the energy flux passing through the cascade.   They emphasize that the conditional dependence they observe is not in violation of the assumptions of the refined Kolmogorov theory since changes in the energy flux should change the small scales.   
Sreenivasan et al. \cite{Sreenivasan:1996} use measurements in the atmospheric boundary layer to demonstrate the conditional dependence of structure functions on the velocity.  They identify this conditional dependence as a result of mixed averages over regions of the flow with different energy dissipation rate and show that when properly normalized by the instantaneous local energy dissipation rate that the conditional dependence is removed in agreement with Kolmogorov's refined similarity hypotheses.
More recently, Mouri et al. \cite{Mouri:2006} explored the effects of large scale fluctuations of the turbulence energy dissipation rate.  They measure grid and boundary layer turbulence and clearly confirm that the large scale energy fluctuations exist and that they affect small-scale statistics.   They explicitly state the the large scale fluctuations do not affect the power law scaling or the coefficients of second order structure functions in the inertial range.   

There is another set of literature exploring time dependent energy input in turbulence that has identified the presence of response maxima when the energy input oscillates with a period on the order of the large eddy turn-over time.  This effect was first predicted in a mean field theory~\cite{vanderHeydt:2003}. It has been explored in a variety of models, numerical simulations and experiments~\cite{Cadot:2003, vanderHeydt:2003, Kuczaj:2006, Bos:2007, Kuczaj:2008, Jin:2008}. However, this work has focused on modulation periods near the turn-over time and seems not to have considered the effects on structure functions, which are most prominent for long modulation periods.

In this paper we present a series of experimental measurements of the effects of time-dependent energy input on the small scales of turbulence.  We focus on second order structure functions where the effects of internal intermittency are small.  We find that the coefficient of the inertial range scaling law depends on the fluctuations in the large scale energy input and measure coefficients that are more than 20\% below the value for the continuously driven case. 

\section{Experiment}

The turbulence is generated in an octagonal Plexiglas tank that is 1 x 1 x 1.5 m$^3$ filled with approximately 1100 $l$ of filtered and degassed water.  Two identical octagonal grids oscillate in phase to generate the turbulence.  The grids have 8 cm mesh size, 36\% solidity, and are evenly spaced from the top and bottom of the tank with a 56.2 cm spacing between grids and a 1 cm gap between the grids and the tank walls.  The grid oscillation has 12 cm amplitude and is powered by an 11kW motor.  In these experiments, the grids were oscillated with frequencies up to 4 Hz which allows Taylor Reynolds numbers up to $R_\lambda=271$.  Details about the experimental setup are available in Blum et al.~\cite{Blum:2010}.

\begin{figure}
\begin{center}
\includegraphics[width=4.5in]{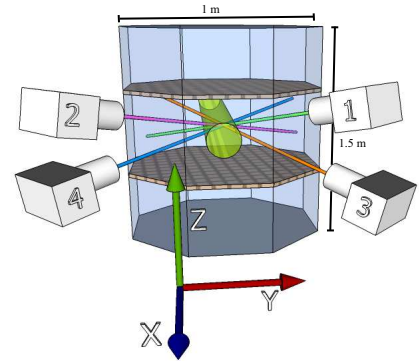}
\caption{\label{fig:cameras} Experimental setup.  Four high speed cameras obtain stereoscopic images of a (5 cm)$^3$ volume at the center of the flow that is illumined by a pulsed Nd:YAG laser with 50 W average power. }
 \end{center}
 \end{figure}

We use stereoscopic particle tracking using four cameras as shown in figure~\ref{fig:cameras}.  The cameras are two Bassler A504K video cameras capable of 1280 x 1024 pixel resolution at 480 frame per second, and two Mikrotron MC1362 cameras with the same pixel resolution and data rates, but with greater sensitivity.  A 5 x 5 x 5 cm$^{3}$ detection volume at the center of the flow was illuminated with a pulsed 50 W Nd:YAG laser.  A real-time image compression circuit with compression factors of 100 to 1000 enables us to acquire data continuously, which allows access to large data sets of particle trajectories~\cite{Chan:2007}.

\begin{figure*}
\begin{center}
\includegraphics[width=5.7in]{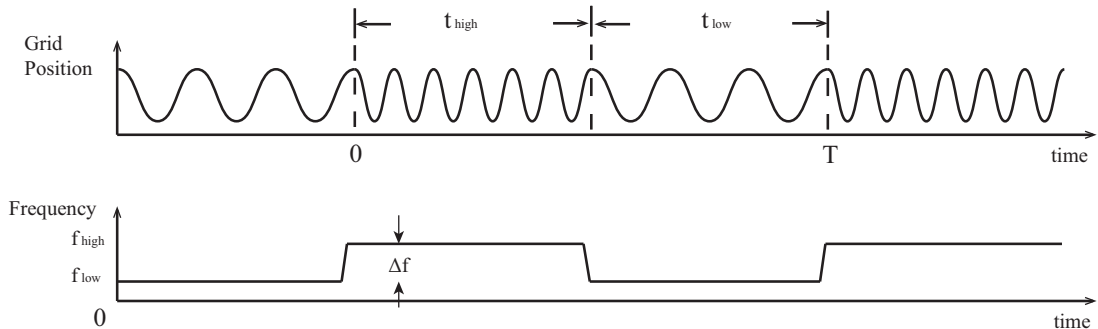}
\caption{\label{fig:cartoon} A sketch of the position and frequency of the oscillating grids as a function of time.  $t_{high}$ is the time over which the grids oscillate at the higher frequency, $t_{low}$ is the time at lower frequency.  $T$ is the cycle period, the time to complete one cycle of modulation from high to low frequency. $f_{high}$ is the high frequency of grids and $f_{low}$ is the low frequency.  $\Delta f$ is the frequency differences between $f_{high}$ and $f_{low}$. }
 \end{center}
 \end{figure*}

\begin{figure*}
\begin{center}
\includegraphics[width=4.5in]{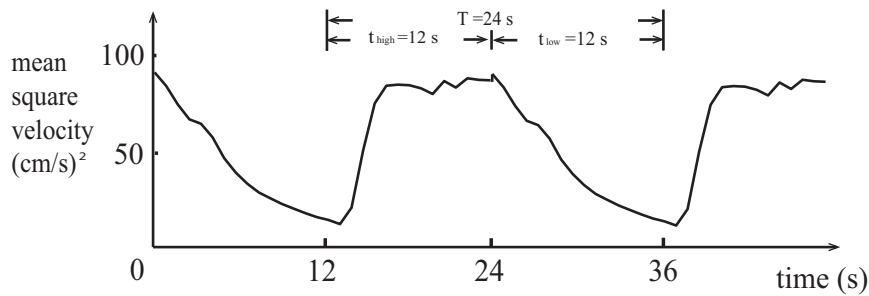}
\caption{\label{fig:cartoon_energy} Time dependence of the mean square velocity measured by phase averaging over many cycles for an experiment with $f_{high} = 3$ Hz, $f_{low} = 0$ Hz, Period $T =$ 24 s, and $50 \%$ duty cycle.  Both the first and second cycle are phase averages over the whole experiments and hence identical.}
 \end{center}
 \end{figure*}

Previous work with this experiment has shown that there are measurable fluctuations in the energy input even when the driving frequency of the oscillating grids is constant~\cite{Blum:2010}.  Here we augment this effect by modulating the driving frequency of the oscillating grids.  
For example, rather than driving the grids continuously at 3 Hz, we can drive it at 3 Hz for 15 s, and then halt for 15 s, and repeat.   This produces a periodic time dependence in the energy input with a longer time scale than the grid oscillation period.    Figure \ref{fig:cartoon} shows a schematic of the frequency modulation along with variable definitions.  In this paper, we explore three different ways to augment the fluctuations in large scale energy input:  (1) change $T$, the time to complete one modulation cycle (2)  change the frequency modulation by holding $f_{high}$ constant and changing $f_{low}$ from 0 up to $f_{high}$, and  (3) changing the duty cycle $t_{high}/T$.   Figure~\ref{fig:cartoon_energy} shows a specific example of the time dependence of the mean square velocity, $\langle u_i u_i \rangle$, which is a measure of the energy in the large scales.  The mean is obtained as a phase average over many cycles.      It takes time for energy to propagate from the grid to the detection volume, so the energy lags several seconds after the grid frequency changes.   

The inertia of the system used to drive the grids limited the rate at which the driving frequency could be changed.  We were able to reduce the time required to stop or start to less than 1/3 of a second  by minimizing the inertia in the experiment.  The original version of this apparatus~\cite{Blum:2010} used a flywheel to improve symmetry between the up and down stroke of the oscillating grids.  For this experiment we replaced the flywheel with a coupler.     For the run with $f_{high} = 3$ Hz shown in figure~\ref{fig:cartoon_energy}, the start time is less than one oscillation and accounts for less than 3\% of the data.  However, limitations from the inertia of the drive system did limit our experiments to periods of $T= 3$ s and greater, which resulted in the period of the modulation of the energy input always being longer than the large scale turn over time.
 
We conducted three sets of experiments to explore the effects of fluctuations of large scale energy input on small scales.   Parameters for each of the experiments are given in table~\ref{table:LSI}.   In the first set of experiments we made measurements with period $T$ of 3, 6, 12, 24, 48, and 384 seconds while always modulating the grid frequency with ($f_{high}$ - $f_{low}$) =  (3 - 0) Hz, with a duty cycle of 50\%.  We will refer to these experiments as ``varying the period''.  In the second set of experiments, we held $f_{high} =$ 3 Hz and made measurements with$f_{low}$ of 3, 2, 1, and 0 Hz to get  ($f_{high}$ - $f_{low}$) = (3 - 3), (3 - 2), (3 - 1), (3 - 0) Hz with $T=$ 30 s period and 50\% duty cycle.  We will refer to these experiments as ``varying the amplitude''.  In the third set of experiments we made measurements with duty cycles of 0\%, 25\%, 50\%, 75\% and 100\%, while always modulating the grid frequency with ($f_{high}$ - $f_{low}$) =  (3 - 0) Hz and a period of $T = 30$ s.  We will refer to these experiments as ``varying the duty cycle''.   We also took data with continuous drive at grid frequencies ranging from 1 Hz to 4 Hz to vary Reynolds number as our control group to show that the effects we observe cannot be simply attributed to the changes in Reynolds number.  

\begin{figure} [h]
\begin{center}
\includegraphics[width=2.8in]{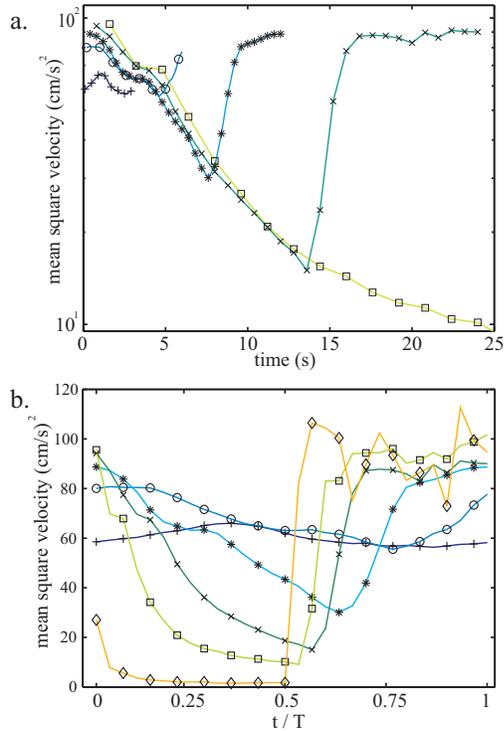}
\caption{\label{fig:energytime} (a) Time dependence of the mean square velocity measured by phase averaging over many cycles.  The motor is halted at t = 0, and turned back on after half a cycle period, $t=T/2$.  Data is from the experiments with varying period.  Symbols represent the cycle period, $T$ of $+$= 3 s, $\circ =$ 6 s, $\ast =$ 12 s, $\times =$ 24 s,  $\Box =$ 48 s.  The symbols $\circ$ and $+$ are only plotted every four data points, and other symbols show every data point.  (b) The fluctuating energy versus $t/T$ with an additional data set $T =$ 384 s ($\diamond$).   Here all data sets have symbols plotted for every other data point.}
 \end{center}
 \end{figure}

In figure \ref{fig:energytime}a we show the time dependence of the mean square velocity, $\langle u_i u_i \rangle$, for the set of experiments varying the period.   Time zero is defined as the time when the energy input halts.  For all of these experiments, the energy dissipates at approximately the same rate, so the decay curves nearly collapse.  After half a period, the energy input resumes.  For the experiments of longer period such as $T =$ 48 s, the energy has decayed to 10\% of its initial value after half a period.  In figure \ref{fig:energytime}b, this data is shown with time normalized by the period.   One additional data set with $T =$ 384 s is added in this plot.    Only for this data set with a very long period does the fluid become approximately quiescent before the energy input is resumed. 

\section{\label{sec: Results}Results}
\subsection{\label{sec:Coefficients_of_scaling_law}Coefficients of inertial range scaling law}

\subsubsection{Varying Period}
\begin{figure} [h!]
\begin{center}
\includegraphics[width=3in]{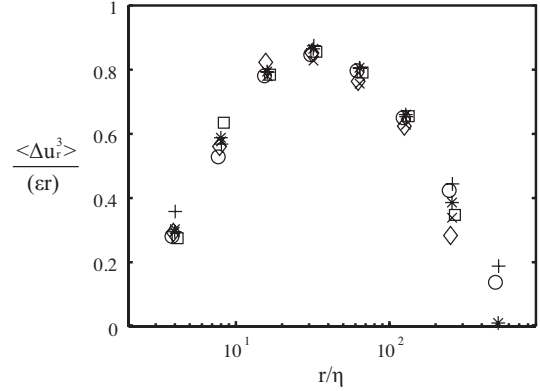}
\caption{\label{fig:structure_3rd} Third order compensated structure functions for the experiments with varying period.   Symbols represent the cycle period, $T$ of $+$= 3 s, $\circ =$ 6 s, $\ast =$ 12 s, $\times =$ 24 s,  $\Box =$ 48 s, $\diamond =$ 384 s.   Driving frequency modulation is ($f_{high}$ - $f_{low}$) =  (3 - 0)Hz  and the duty cycle is $50\%$. }
 \end{center}
 \end{figure}

Figure~\ref{fig:structure_3rd} shows the third order structure functions of the experiments varying the period.   The energy dissipation rate is determined from this data and the four-fifths law.  When compensated by $\varepsilon r$, the inertial and dissipation ranges of these third order structure functions collapse fairly well, suggesting that the small scales of these turbulent flows are similar.   
  
\begin{figure}[h!]
\begin{center}
\includegraphics[width=3in]{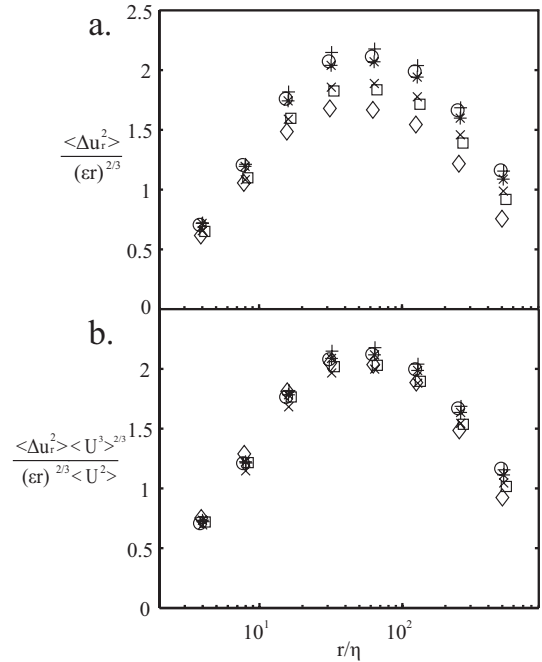}
\caption{\label{fig:structure} (a) Second order compensated structure functions for the experiments with varying period.  Symbols are the same as \protect{figure~\ref{fig:structure_3rd}}.   (b) Second order structure functions scaled by the ratio of moments of the energy dissipation rate predicted by the refined model in equation~\protect{\ref{refined_model}}. }
 \end{center}
 \end{figure}

  However, the compensated second order structure functions shown in figure~\ref{fig:structure}a do not collapse well at all.   The maximum of these compensated structure functions, which is an estimate of the coefficient in the inertial range scaling law, shows a 20\% decrease as the period increases.  Increasing the fluctuations in the energy input does have a significant effect on the small scales of the flow.   The shape of the second order structure functions shows little change, which is consistent with the idea that fluctuations in the energy input at large scales primarily change the coefficients in scaling laws while leaving the scaling exponents unchanged.   Figure~\ref{fig:structure}b shows the second order structure functions scaled by the prediction of equation~(\ref{refined_model}).  The good collapse of these curves after scaling indicates the effects of fluctuations in the energy input are largely captured by the refined model.

\begin{figure}[h!]
\begin{center}
\includegraphics[width=3in]{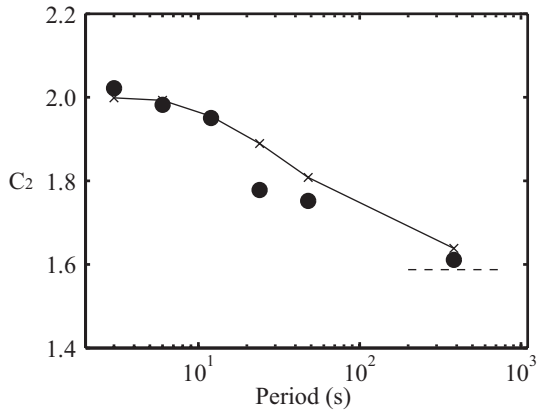}
\caption{\label{fig:category_period_gamma} Experimental measurements of inertial range scaling coefficient ($\bullet$) along with the prediction of  the refined model ($\times$) for the experiments of varying period.  The dotted line represents the prediction of the model by Monin and  Yaglom. }
 \end{center}
 \end{figure}

Figure~\ref{fig:category_period_gamma} shows the measured coefficient of the inertial range scaling of the second order structure function, commonly labelled as Kolmogorov constant $C_2$.  The decrease in the `constant' as the period increases is a clear indication that the previous assessment by Mouri et al.~\cite{Mouri:2006} and Praskovsky et al. \cite{Praskovsky:1993} that large scale fluctuations do not affect second order structure functions is only an approximation that is valid in cases where the fluctuations in the energy input are small.    Figure~\ref{fig:category_period_gamma} also shows the prediction of our refined model from equation~(\ref{refined_model}) with the model value of $C_2=2.0$.  The experimental measurement and the refined model are in fairly good agreement.   There are many possible factors that contribute to the difference between the measurements and the model, including the difficulty in measuring scaling coefficients at modest Reynolds number and limitations of the estimate $\varepsilon \propto U^3/L$ in equation~(\ref{refined_model_u}).  The dotted line is the prediction of the model by Monin and Yaglom. Our experimental measurements of the inertial range coefficient approaches this dotted line when the period is long as it should since in that case we are approaching the situation Monin and Yaglom consider where the energy input is constant in time for both the low frequency and high frequency state. 

 Measuring scaling coefficients from this data at modest Reynolds numbers has some difficulties.   From the third order structure functions we extracted the energy dissipation rate by averaging the three bins at the maxima between $r/\eta=$15 and 68.  For the second order structure functions, we used this same definition of the inertial range even though the peak of the second order compensated structure functions are at slightly larger $r$.   This results in measured second order scaling coefficients being below the peak value.   We tried using a different inertial range for the second order data.  This makes small changes in the magnitude of the scaling coefficients, but has no effect on the conclusions we draw.  Data at larger Reynolds numbers will be necessary to provide more precise quantitative measurements of how scaling coefficients depend on fluctuations in the energy input.

\begin{figure} [h!]
\begin{center}
\includegraphics[width=3in]{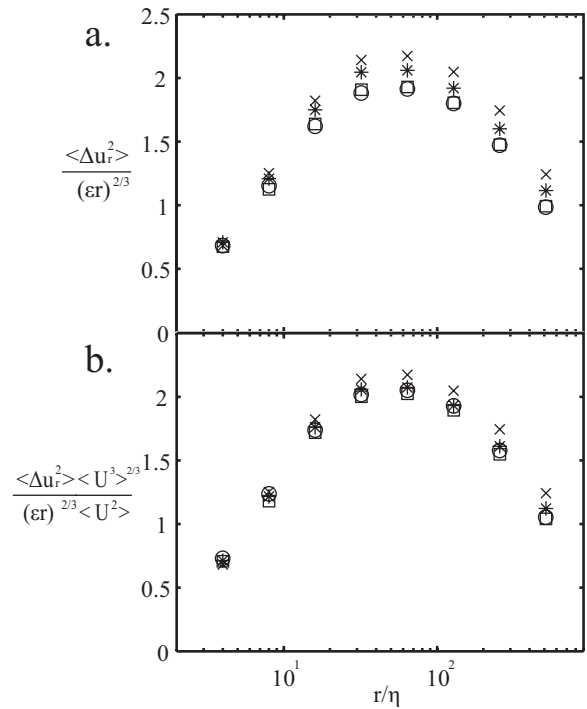}
\caption{\label{fig:structure_3Hz} (a) Second order compensated structure functions for the experiments with varying amplitude.   (b) Second order structure functions scaled by the ratio of moments of the energy dissipation rate predicted by the refined model.  Symbols represent different frequency modulations of  ($f_{high}$ - $f_{low}$). $\times =$ (3 - 3) Hz, $\ast=$ (3 - 2) Hz, $\Box =$ (3 - 1) Hz, $\circ =$ (3 - 0) Hz.  Cycle period $T$ is 30 s, and the duty cycle is $50\%$. }
 \end{center}
 \end{figure}
 
\subsubsection{Varying Amplitude}
Similar effects of the large scale energy fluctuations on small scales are also seen in the experiments where amplitude of the energy input is varied by changing the grid oscillation frequency.  Figure \ref{fig:structure_3Hz}a shows the second order compensated structure functions  for the data sets with varying amplitude. Similar to the experiments varying the period, the curves do not collapse, indicating that the coefficient of the scaling law depends on the large scales.  Figure~\ref{fig:structure_3Hz}b shows the second order structure functions scaled by the prediction of equation~(\ref{refined_model}).  The better collapse of these curves after scaling again indicates that the refined model is accurately describing the effects of fluctuating energy input.   

\begin{figure}
\begin{center}
\includegraphics[width=3in]{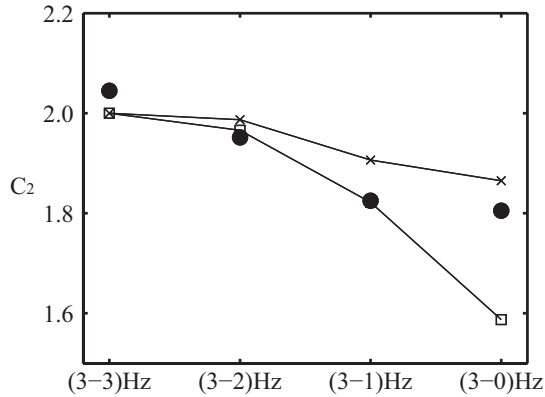}
\caption{\label{fig:category_3Hz_gamma} 
Experimental measurements of the inertial range scaling coefficient ( $\bullet$), compared with predictions from the refined model ($\times$), and the Monin and Yaglom model ($\Box$) for the experiments of varying amplitude.  The predictions of the refined model and the Monin and Yaglom model assume $C_2$ = 2. }
 \end{center}
 \end{figure}

Figure~\ref{fig:category_3Hz_gamma} shows the measured coefficient of the inertial range scaling along with predictions from the refined model and the Monin and Yaglom model.  The main point is that increasing the amplitude of the fluctuations in the energy input systematically decreases the constant as predicted.   Quantitatively, the refined model has coefficients larger than those measured meaning that it underestimates the effect of the large scale fluctuations.  This deviation is likely due to the refined model using the time dependence of the rms velocity to estimate the fluctuations in the energy input, which does not capture all of the fluctuations.  The Monin and Yaglom model works well for small amplitude of the energy input fluctuations, but for the largest fluctuation amplitude (3-0)Hz, it predicts a much larger effect of the large scale fluctuations than are observed experimentally.  This is expected since this data is for period $T=30$ s, and there is not enough time for the energy to decay to the constant values assumed by the Monin and Yaglom model.

For the experiments varying the amplitude of the fluctuations in the energy input, we did not directly measure the phase averaged fluctuating velocity needed in the refined model.  To make predictions with this model, we had to model the fluctuation velocity using the known values for continuous driving at different frequencies and the decay rate data in figure~\ref{fig:energytime}.  The limitations of this model likely also contributes to the poorer agreement with the refined model in this case.

\begin{figure}
\begin{center}
\includegraphics[width=2.8in]{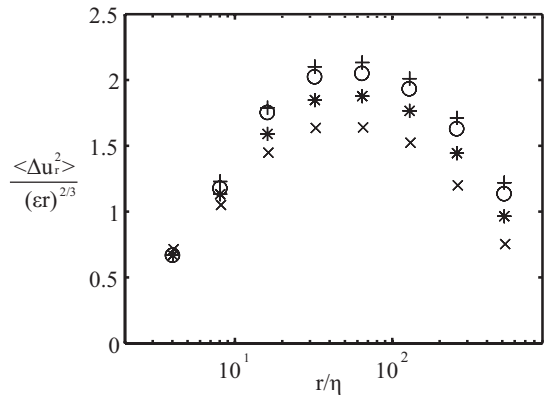}
\caption{\label{fig:structure_dutycycle} Second order compensated structure functions for the experiments with varying duty cycle. Symbols represent the duty cycle of $+$= 100\%, $\circ =$ 75\%, $\ast =$ 50\%, $\times =$ 25\%,  $\Box =$ 48 s, $\diamond =$ 384 s.   Driving frequency modulations is ($f_{high}$ - $f_{low}$) =  (3 - 0)Hz  and the period $T$ is $30$ s.  }
 \end{center}
 \end{figure}

\begin{figure}
\begin{center}
\includegraphics[width=2.8in]{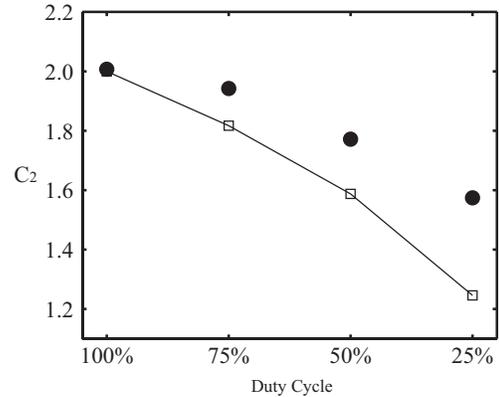}
\caption{\label{fig:category_dutycycle_gamma} Experimental measurements of the inertial range scaling coefficient ($\bullet$), and the prediction of the Monin and Yaglom model ($\Box$) for the experiments of varying duty cycle.  }
 \end{center}
 \end{figure}
 
 \subsubsection{Varying Duty Cycle}
The set of experiments varying the duty cycle in figure~\ref{fig:structure_dutycycle}  also shows that the compensated second order structure functions show strong dependence on fluctuations in the energy input.      We show the measured  inertial range scaling coefficient in figure~\ref{fig:category_dutycycle_gamma}.  When the duty cycle is smaller, we observe a smaller coefficient.  For 25\% duty cycle we see the smallest value of the inertial range scaling coefficient of 1.58.    Note that 25\% duty cycle and 75\% duty cycle do not have the same coefficient.  Because times with large energy input dominate the moments of the energy dissipation rate, the effects on the coefficient are largest for low duty cycle where bursts of large energy input are followed by a long quiescent period.  
 
The predictions of the Monin and Yaglom model shown in figure~\ref{fig:category_dutycycle_gamma} are consistently below the measured coefficients.  We expect that if the experiments were performed for larger period rather than $T=30$ s they would approach the Monin and Yaglom predictions.

 \subsubsection{Varying Reynold's number}
\begin{figure}[h!]
\begin{center}
\includegraphics[width=3.2in]{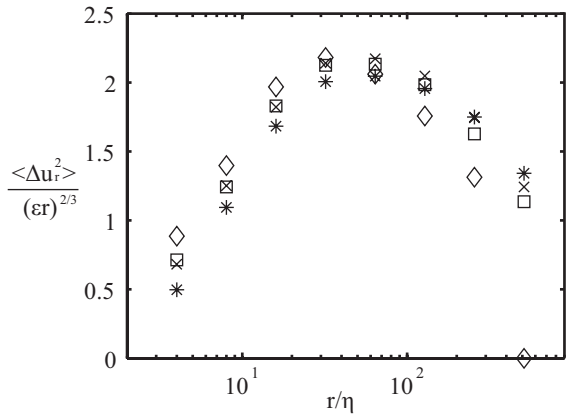}
\caption{\label{fig:structure_Reynold} Second order compensated structure functions for the experiments with varying Reynolds number.   Symbols represent different Reynolds number, $R_\lambda$ of $\ast =$ 271, $\times =$ 250,  $\Box =$ 237, $\diamond =$ 163. }
 \end{center}
 \end{figure}

The set of experiments varying Reynolds number for constant energy input in figure~\ref{fig:structure_Reynold} shows that the inertial range scaling coefficients for the second order structure functions do not have strong dependence on Reynolds number.  We vary Reynolds number from $R_\lambda=$139 at 1 Hz continuous driving to $R_\lambda=$ 271 at 4 Hz continuous driving.  The shape of the structure function changes at the lowest Reynolds number as expected, but after using the third order structure functions to determine the energy dissipation rate, the peak value remains relatively constant.  This confirms that the variation we observe in the Kolmogorov constant is not simply the result of different effective Reynolds numbers in different experiments. 

\begin{table*} 
\begin{center}
\begin{tabular}{|p{1.8cm} |p{0.9cm} |p{0.7cm} |p{1cm} |p{1cm}|p{1.1cm} |p{0.9cm}|l|p{1.4cm} |l|}
\hline
  & $f_{high}$ (Hz)& $f_{low}$ (Hz) & $T$ (s) & Duty Cycle &U (cm/s) & $L$ (cm) & $\tau$ (s) & $\varepsilon$ (cm$^2$/s$^3$) & $R_\lambda$ \\ \hline
\multirow{2}{*}{Varying } &3 & 3 & 30 & 50$\%$ & 5.46 & 7.69 & 1.41 & 21.2 & 250 \\  \cline{2-10}
 \multirow{2}{*}{amplitude} &3 & 2 & 30 & 50$\%$  & 4.72 & 7.48 & 1.58 & 14.1 & 230 \\  \cline{2-10}
&3 & 1 & 30 & 50$\%$  & 4.23 & 7.13 & 1.69 & 11 & 213 \\  \cline{2-10}
 &3 & 0 & 30 & 50$\%$  & 4.21 & 7.16 & 1.7 & 10.4 & 212 \\ \hline

 &3 & 0 &  3 & 50$\%$  & 4.44 & 8.28 & 1.86 & 10.6 & 235 \\ \cline{2-10}
\multirow{3}{*}{Varying}&3 & 0 &  6 & 50$\%$ & 4.71 & 8.51 & 1.81 & 12.3 & 245 \\ \cline{2-10}
\multirow{3}{*}{period}&3 & 0 &  12 & 50$\%$ & 4.54 & 8.44 & 1.86 & 11.1 & 240 \\ \cline{2-10}
&3 & 0 & 24 &  50$\%$ & 4.42 & 7.87 & 1.78 & 11 & 228 \\ \cline{2-10}
&3 & 0 & 48 & 50$\%$ & 4.07 & 6.48 & 1.59 & 10.4 & 198 \\ \cline{2-10}
&3 & 0 &  384 & 50$\%$ & 4.06 & 5.76 & 1.42 & 11.6 & 187 \\ \hline

 \multirow{2}{*}{Varying} &3 & 0 & 30 & 100$\%$ & 5.46 & 7.69 & 1.41 & 21.2 & 250 \\ \cline{2-10}
\multirow{2}{*}{duty cycle}&3 & 0 &  30 & 75$\%$ & 4.92 & 7.58 & 1.54 & 15.7 & 236 \\ \cline{2-10}
&3 & 0 & 30 & 50$\%$  & 4.21 & 7.16 & 1.7 & 10.4 & 212 \\ \cline{2-10}
&3 & 0 &  30 & 25$\%$ & 3.27 & 6.86 & 2.1 & 5.1 & 183 \\ \hline

 \multirow{2}{*}{Varying} &1 & N/A & N/A& 100$\%$ & 1.96 & 6.59 & 3.36 & 1.15 & 139 \\ \cline{2-10}
\multirow{2}{*}{Reynolds}&2 & N/A & N/A& 100$\%$ & 4.05 & 9.3 & 2.3 & 7.15 & 237 \\ \cline{2-10}
\multirow{2}{*}{Number}&3 & N/A & N/A& 100$\%$ & 5.46 & 7.69 & 1.41 & 21.2 & 250 \\ \cline{2-10}
&4 & N/A & N/A& 100$\%$ & 7.14 & 6.87 & 0.96 & 52.9 & 271 \\ \hline

\end{tabular}
\caption{ Experimental parameters and resulting statistics for different sets of experiments.  Note that the case of $f_{high}$ = 3 Hz, $f_{low}$ = 3 Hz, Duty Cycle 50\% is the same data as the case of $f_{high}$ = 3 Hz, $f_{low}$ = 0 Hz, Duty Cycle 100\%.}
\label{table:LSI} 
\end{center}

\end{table*}

\subsection{\label{sec: Conditional Structure Function} Conditional Structure Functions}

Previous work has used conditional structure functions to quantify the effects of the large scales on small scales in turbulent flows~\citep{Praskovsky:1993, Sreenivasan:1996, Sreenivasan:1998, Blum:2010, Blum:2011}.   Velocity differences between two points separated by $\textit{\textbf{r}}$ are dominated by structures near scale $\textit{\textbf{r}}$ while velocity sums of two points are dominated by the large scales in the flow. So moments of velocity differences conditioned on sums provide a convenient way to observe the effects of the largest scales on other scales.  We find that conditional structure functions provide a more sensitive measurement of the existence of fluctuations in the large scale energy input than the coefficients of inertial range structure functions.  However, theoretical tools to predict the effects of large scale fluctuations on conditional structure functions are not available.   In this section we present measured conditional structure functions as we systematically change the fluctuations in the energy input.  

\begin{figure}
\begin{center}
\includegraphics[width=2.4in]{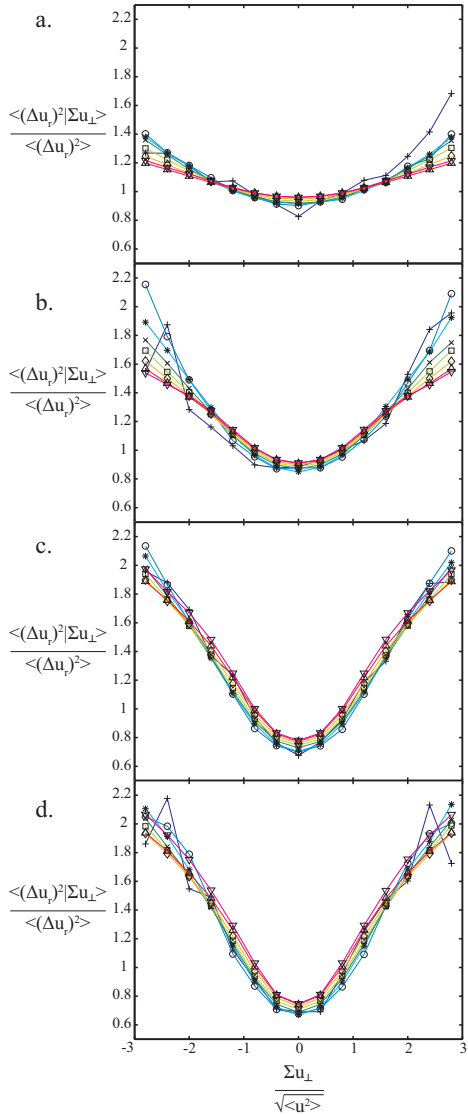}
\caption{\label{fig:3Hz_all} Eulerian second order conditional structure function versus large scale velocity for the experiments with varying amplitude. The frequencies modulated were  ($f_{high}$ - $f_{low}$) =  (a) (3 - 3) Hz (b) (3 - 2) Hz, (c) (3 - 1) Hz, (d) (3 - 0) Hz.  Each curve represents the following separation distances $r/\eta$: $+$ = 2.67 to 5.33, $\circ$ = 5.33 to 10.67, $\ast$ = 10.67 to 21.33, $\times$ = 21.33 to 42.67, $\Box$ = 42.67 to 85.33, $\diamond$ = 85.33 to 170.67, $\vartriangle$ = 170.67 to 341.33, $\triangledown$ = 341.33 to 682.67.}
 \end{center}
 \end{figure}

\subsubsection{Varying amplitude}

Figure~\ref{fig:3Hz_all} shows conditional structure functions for the data sets varying the amplitude of the fluctuations in the energy input.   We condition the structure function on the velocity component that is transverse to $\textit{\textbf{r}}$ denoted $\Sigma u_\perp$.  In order to compare the conditional structure function for different length scales, we normalize the vertical axis by the unconditioned structure function.   The horizontal axis is normalized by the characteristic velocity $U=(\langle u_i u_i/3 \rangle)^{1/2}$.   In figure~\ref{fig:3Hz_all}a for constant driving of the oscillating grids, we see the results published by Blum et al.~\cite{Blum:2010} that the conditional structure functions for all length scales show a similar dependence on the large scale velocity.  There is a slight dependence on length scale with the smallest length scales showing a stronger dependence on the large scale velocity.  This small dependence on length scale remains unexplained since it is the opposite of the expectation that the small scales are approaching universality.  The same effect is seen in DNS data in Ref.~\cite{Blum:2011}.    However, in this paper we are focusing on fluctuations of the energy input and we will see that these produce much bigger effects than the small differences for different length scales.

\begin{figure}[h!]
\begin{center}
\includegraphics[width=3.0in]{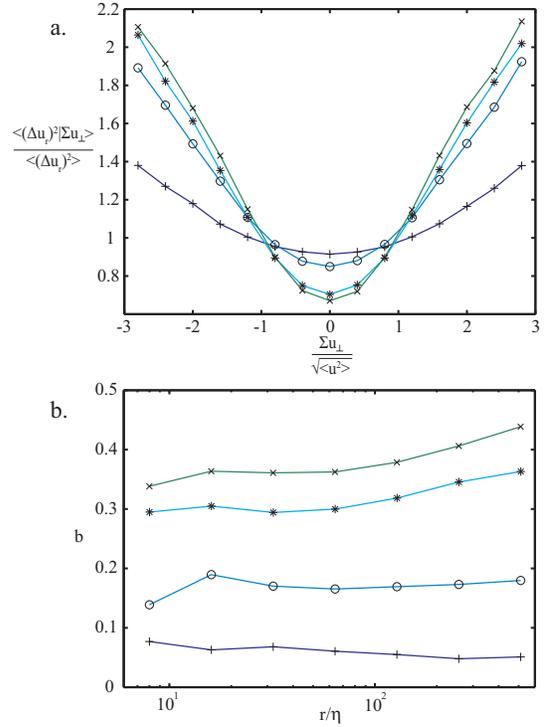}
\caption{\label{fig:3Hz}(a) The velocity dependence of conditional second order structure functions of one separation distance $r/\eta =$ 10.67 to 21.33 for the experiments with varying amplitude.  ($f_{high}$ - $f_{low}$) = (3 - 3) Hz ($+$), (3 - 2)Hz ($\circ $), (3 - 1) Hz ($\ast $), (3 - 0) Hz ($\times$) with 50\% duty cycle and $T =$ 30 s  (b) The coefficient $b$ as a function of the separation distance for the experiments with varying amplitude.  Symbols are the same as (a). }
 \end{center}
 \end{figure}

Figures~\ref{fig:3Hz_all}b-d show that increasing the fluctuations in the energy input produces a large increase in the dependence of the conditional structure functions on the large scale velocity.    In each sub-figure, the curves for different length scales remain very similar, which confirms the fact observed earlier that fluctuating energy input does not change the length scale dependence.  It primarily changes a pre-factor scaling the entire structure function.   Note that figure~\ref{fig:3Hz_all}a still has dependence on the large scale velocity even though the oscillating grid is driven at a constant 3 Hz frequency.  We interpret this as fluctuations in the energy input that remain even in the case of constant driving~\citep{Blum:2010}. To more directly compare the effects of changing the energy input fluctuations, we extract the curve for $r/\eta$ = 10.7 to 21.3 from figure~\ref{fig:3Hz_all}(a, b, c and d) and plot them on one graph as shown in figure \ref{fig:3Hz}a.  In figure \ref{fig:3Hz_all} and figure \ref{fig:3Hz} the symmetry around zero large scale velocity is a result of conditioning on the transverse component of the large scale velocity for which $\Sigma u_\perp > 0$ is indistinguishable from $\Sigma u_\perp < 0$.

To quantify the observed dependence of the conditional structure function, we fit all the curves in figure \ref{fig:3Hz_all} to the functional form $au^4+bu^2+c$.  Figure~\ref{fig:3Hz}b shows the fit coefficient  $b$ as a function of the separation distance $r/\eta$.   The coefficient $b$ measures the curvature of the conditional structure functions at the origin, and it captures the primary dependence on the large scale velocity.    Measuring the coefficient of the second order term $b$ is also keeping with a previous study ~\citep{Sreenivasan:1998}.  There is an increase by more than a factor of 5 in the curvature, $b$,  as the fluctuations in the energy input increase from driving at 3 Hz continuously to alternating between 3 and 0 Hz.  The degree to which all length scales show similar dependence on the large scales can also be evaluated from figure \ref{fig:3Hz}b.  In section \ref{sec: Summary} we will show that changes in $b$ are closely related to the changes in the inertial range scaling coefficient that we presented in section \ref{sec:Coefficients_of_scaling_law}.

\subsubsection{\label{sec:Varying period} Varying period}

\begin{figure}
\begin{center} 
\includegraphics[width=2.8in]{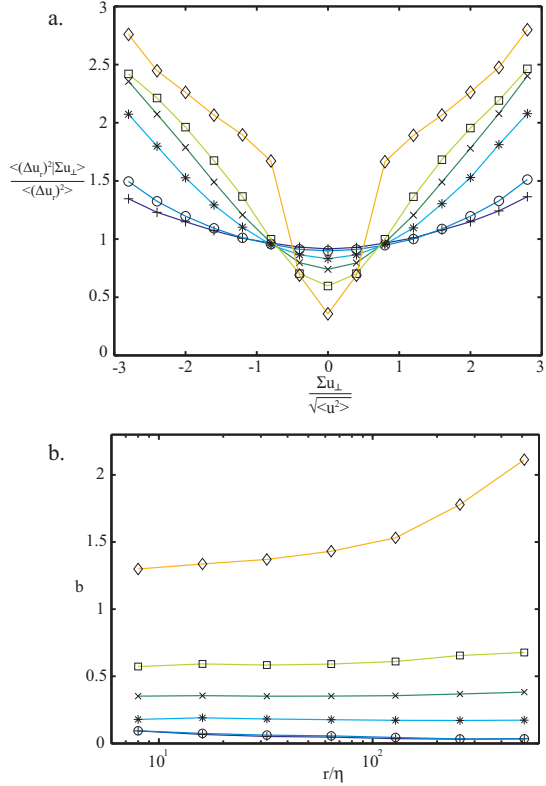}
\caption{\label{fig:ontime} (a) The velocity dependence of second order conditional structure functions of one separation distance $r/\eta =$ 10.67 to 21.33 for the experiments with varying period.  Symbols represent the cycle period, $T$ of $+ =$ 3 s, $\circ =$ 6 s, $\ast =$ 12 s, $\times =$ 24 s,  $\Box =$ 48 s, $\diamond =$ 384 s.   Driving frequency modulations is ($f_{high}$ - $f_{low}$) =  (3 - 0)Hz, and the duty cycle is $50\%$. (b) The coefficient $b$ as a function of separation distance for the experiments with varying period.  Symbols are the same as (a). }
 \end{center}
 \end{figure}

Figure \ref{fig:ontime}a shows the conditional second order structure functions  for the experiments with varying period.  When the period $T$ increases, there is a stronger dependence on large scale velocity.   The two shortest periods $T =$ 3 s and 6 s have similar and relatively low curvatures.  Increasing the period allows the turbulence to decay closer to quiescent before the energy input resumes, so the conditional dependence on the large scale velocity is stronger at longer periods.   For the very long period, $T=384$ s, the conditional structure function has a different shape with a sharp minimum at the center of a region with less curvature.    This is the result of the high energy state providing the samples with large velocity sum, while the low energy state provides only samples with velocity sum near zero.  For this data at  $T=384$ s there is also a much stronger dependence on the length scale as shown in figure \ref{fig:ontime}b.

\begin{figure}
\begin{center}
\includegraphics[width=2.8in]{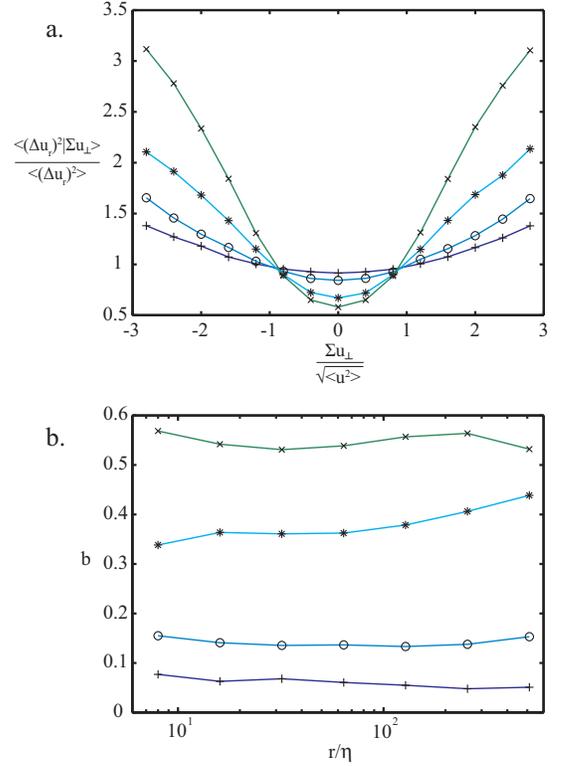}
\caption{\label{fig:duty_cycle}  (a) The velocity dependence of conditional second order structure functions of one separation distance $r/\eta =$ 10.67 to 21.33 for the experiments with varying duty cycle.  Each curve shows the duty cycle of $+ =$ 100\% , $\circ = $ 75\% , $\ast = $ 50\%, $\times = $ 25\% with driving frequency modulations ($f_{high}$ - $f_{low}$) =  (3 - 0)Hz and period $T = $30 s. (b) The coefficient $b$ as a function of separation distance for the experiments with varying duty cycle.  Symbols are the same as (a).}
\end{center}
\end{figure}  
\subsubsection{\label{sec:Varying Duty Cycle} Varying duty cycle}

Figure \ref{fig:duty_cycle}a shows the second order conditional structure functions for the experiments with varying duty cycle. It shows that reducing the duty cycle produces a large increase in the dependence of the conditional structure functions on the large scale velocity.  The result is consistent with our previous findings that increasing the fluctuations of the large scale energy input increases the dependence of the second order conditional structure functions on the large scale velocity.  Here all length scales show fairly similar dependence on the large scales as seen in figure \ref{fig:duty_cycle}b.

\begin{figure}
\begin{center}
\includegraphics[width=2.2in]{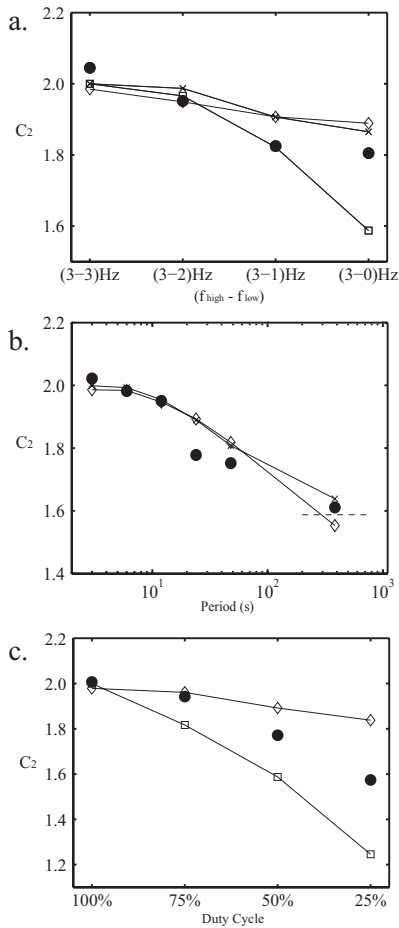}
\caption{\label{fig:cur} The relationship of the curvature $b$ of the conditional second order structure function with the coefficient of the inertial range scaling law.  $\diamond$ is the parameterization 2(1-0.15$b$).  $\bullet$ is the experimental measurements of the inertial range scaling coefficient. $\times$ is the refined model, and $\Box$ is the Monin and Yaglom model.  }
 \end{center}
 \end{figure}
\subsection{\label{sec: Summary}Connecting Conditional Structure Functions and Coefficients of Inertial Range Scaling Law}

The curvature $b$ of the conditional structure functions increases as the fluctuations of the large scale energy input increases.  This suggests that it might be possible to connect $b$ with changes in the coefficients of inertial range scaling law presented in section~\ref{sec:Coefficients_of_scaling_law}.  

A simple linear parameterization $C_2=2(1- 0.15b)$ seems to match the measured scaling coefficients fairly well as shown in figure~\ref{fig:cur}.   However, we do not have a solid theoretical foundation for choosing this functional form and the value of 0.15 is a rough fit.    For weak fluctuations in the energy input, which includes most turbulent flows of interest, this parameterization seems to work fairly well.  But for extreme cases it fails.  At low duty cycles in figure~\ref{fig:cur}c, this parameterization is well above the measured coefficient.  In the limit where one of the states is actually quiescent ($\gamma$ =1 in figure~\ref{fig:contour}), the curvature $b$ should go to infinity while the coefficient of the scaling law would not go negative.   Conditional structure functions and coefficients of inertial range scaling law are both modified by fluctuations in the large scale energy input of turbulence.   A more complete understanding of the relationship between these two could be very useful, since the effects of fluctuations in the large scale energy input are much easier to measure using conditional structure functions.

\section{Discussion}
In this paper we have focused on inertial range effects of fluctuations in the energy input because they are most easily measured with our apparatus.  But it should be noted that the non-universality of the inertial range scaling coefficients implies non-universality of the functional form of structure functions in the dissipation range.   Because the Kolmogorov scale depends on the energy flux, the functional form in the dissipation range will depend on the distribution of the energy flux which depends on the fluctuations of the energy input at large scales.  

A problem facing research into the effects of large scale fluctuations on the small scales of turbulence is that the terminology that has accumulated over many years is not always as clear as it could be.  The word `intermittency' appears to have entered the turbulence literature to describe the fluctuations between turbulent and non-turbulent fluid flowing past a point in a free shear flow.  For example, the textbook by Hinze in 1959 uses `intermittent' only in this sense.  The second edition of this textbook in 1975 introduces the use of a flatness factor to measure the `degree of intermittency' (p. 242), but even here, the goal is to quantify the fraction of the time that turbulence occurs.  Over the decades a major change has occurred in how the word intermittency is used.   In the parlance of a large part of the turbulence research community, intermittency has  become associated with the rare events of large dissipation that are responsible for anomalous scaling~\citep{Sreenivasan:1997}.   A good example of this usage is the book by Frisch ~\cite{Frisch:1995} which uses the word `intermittency'  to refer to the fluctuations produced by uneven energy transfer through the cascade which we refer to above as internal intermittency.  He briefly describes the turbulent to non-turbulent fluctuations seen in free shear flows with a footnote that says ``This phenomenon is known as `external intermittency'; its relation to the intermittency discussed in Chapter 8 is not clear''.    In general use, the word `intermittency' has often taken on a connotation about large deviations from the mean that is entirely absent in the standard English definition of the word or in the traditional application of this word to turbulent flows.    However, the old terminology is also still used.  In the textbook by Pope (2000), the word intermittency is reserved for the turbulent to non-turbulent fluctuations in free shear flows while small scale effects are called `internal intermittency'.  Other sources  use the phrase `large-scale intermittency' to refer to the turbulent to non-turbulent fluctuations in free shear flows~\citep{Mi:2001}.  

In this paper, we quantify the effects that fluctuations in the energy input at large scales have on the coefficients of inertial range power laws.  The success of models based on the refined similarity hypotheses suggests we should use terminology that connects this phenomenon with the closely related phenomenon of internal intermittency that is already widely understood.  However, the history of the terminology for these phenomena makes it difficult to find suitable terms.  Davidson \cite{Davidson:2004}  provides a clear description of the phenomenon of fluctuations at large scales and uses the phrases `integral-scale intermittency' and `large-scale intermittency' to refer to them in his section 6.5.1.   We prefer this terminology, but the possibility of confusion with the older use of the phrase `large-scale intermittency' led us not to use this terminology in this paper.

One way to view the contributions from this and a previous sequence of papers~\citep{Blum:2010,Blum:2011} is that in quantifying the effects of large scale fluctuations on small scales, we find that large scale fluctuations which affect the entire cascade are a standard feature of turbulence and not a special feature of free shear flows or periodically modulated flows.   Conditional structure functions are a sensitive way to quantify this dependence, and with them we find that the effects of large scale fluctuations can be detected in all flows except for a few special cases like turbulence behind a passive grid~\citep{Blum:2011}.  This observation is in contrast to the usual assessment (see for example Praskovsky et al. \cite{Praskovsky:1993} and Mouri et al.~\cite{Mouri:2006}) where the large scale fluctuations are viewed as not affecting second order statistics except in free shear flows where conditional sampling of the turbulent regime can be used to restore the universal result.   

Our interpretation is that, in general, turbulent flows have fluctuations in the large scale energy input.  In many cases these are not large enough to have measurable effects on second order statistics, but by explicit control of the time dependence of the energy input we can make these effects big enough to produce a 20\% change in the Kolmogorov constant for the second order structure function.    In other flows that appear to have constant energy input such as boundary layers,  von-Karman flow between counter-rotating disks, etc., the strong inhomogeneity allows fluctuations at the large scales to intermittently transport fluid from different parts of the flow creating fluctuations in the energy input rate which should change the constants in inertial range scaling law in ways predicted by equation~(\ref{refined_avg}).  The effects of turbulent to non-turbulent fluctuations in free shear flows are then seen to be a special case of this more general problem of transport in an inhomogeneous flow by the large scale fluctuations.  To be sure it is an extreme case, where the entrained fluid has no vorticity and the viscous super-layer separating turbulent from non-turbulent fluid can be very thin.  But the extreme case is smoothly connected to other flows where the large scale fluctuations entrain fluid with different turbulence characteristics.   For example, experiments in a shearless mixing layer~\citep{Veeravalli:1989, Kang:2008} can continuously vary the turbulence on the two sides of the mixing layer from the extreme case of turbulent/non-turbulent to the case where the turbulence on both sides of the layer are the same.  

In the future, we hope that the community can adopt some terminology that will allow us to talk more clearly about fluctuations at the large scales of turbulence.  We have shown here that we can quantify and predict the the effects of large scale fluctuations using a refined similarity framework.  These large scale fluctuations destroy universality in the Kolmorogov 1941 sense in exactly the way that Landau predicted, and they seem to naturally be called `large scale intermittency' since they are to the large scales what internal intermittency is to inertial and dissipation range scales.  Furthermore they are the general case under which the traditional use of the phrase `large scale intermittency' can cleanly fall.     We hope that further work on this topic will develop tools to more precisely quantify the fluctuations  at large scales, and that this will lead to a consensus about the terminology to use in discussing these effects.

 \section{\label{Conclusion}Conclusion}
 
 Previous research has established that the small scales in turbulence are not entirely independent of the large scales~\citep{Kolmogorov:1962, Mouri:2006, Blum:2010}.  Landau's footnote remark suggests that the fluctuations in the energy dissipation due to non-universal large scales will destroy the universality of small scales.  Kolmogorov's paper on the refined similarity hypotheses~\citep{Kolmogorov:1962} identifies that the coefficients of inertial range scaling laws will not be universal.  However, during the extensive effort to understand internal intermittency, the effects of fluctuations in the large scales have been largely ignored.  The consensus in the literature has been that the coefficient of the inertial range scaling law for second order structure functions, known as the Kolmogorov constant $C_2$, is a universal constant~\citep{Praskovsky:1993,Sreenivasan:1995,Yeung:1997,Mouri:2006}.

In this paper, we systematically change the fluctuations in the energy input at the large scales and find that this leads to a decrease in the inertial range scaling coefficient that can be more than 20\%.  An extension of the ideas in Kolmogorov's refined theory provides a model that successfully predicts these changes of the coefficients in inertial range scaling laws.   

We also use structure functions conditioned on the velocity sum to measure the effect of fluctuations of large scale energy input on small scales.  These conditional structure functions are able to identify the effects of fluctuations of the energy input even when the fluctuations are small.  The curvature of the second order conditional structure functions appears to be determined by fluctuations in the energy input in a way similar to the changes in the Kolmogorov constant, but a quantitative understanding of this relationship is not available.

The turbulent flows that have been the focus of most laboratory and simulation work appear to have small enough fluctuations in the energy input that the effects on the second order Kolmogorov constant are usually negligible.      However, in many geophysical flows such as turbulent clouds, the large scale fluctuations are a dominant feature of the flow.   Our measurements show that fluctuations in the energy input at large scales can be determined by measuring the coefficients of inertial range scaling laws for conditional structure functions.  This allows small scale measurements to provide a useful diagnostic of large scale dynamics.   When it is possible to make direct measurements or predictions of the fluctuations in the large scale energy input, then the models we use here can provide prediction of the inertial range scaling coefficients from the properties of the large scales. 

\noindent We would like to acknowledge support from the National Science Foundation under grant DMR-0547712 and DMR-1208990 and from COST Action MP0806.    We thank Susantha Wijesinghe for his expertise on the real-time image compression circuit, and Greg Bewley, Eberhard Bodenschatz, Nick Ouellette, Arkady Tsinober, Zellman Warhaft and Haitao Xu for helpful conversations.



\end{document}